# A Simulation-Optimization Technique for Service Level Analysis in Conjunction with Reorder Point Estimation and Lead-Time consideration: A Case Study in Sea Port

Mohammad Arani*
Department of Systems Engineering
The University of Arkansas at Little Rock
Little Rock, the USA
mxarani@ualr.edu

Saeed Abdolmaleki
Department of Industrial Management
The University of Shahid Beheshti
Tehran, Iran
saeedmaleki31@gmail.com

Maryam Maleki
Department of Systems Engineering
The University of Arkansas at Little Rock
Little Rock, the USA
mmaleki@ualr.edu

Mohsen Momenitabar
Department of Transportations and Logistics
The North Dakota State University
Fargo, the USA
mohsen.momenitabar@ndsu.edu

Xian Liu
Department of Systems Engineering
The University of Arkansas at Little Rock
Little Rock, the USA
xxliu@ualr.edu

*Abstract*—This study offers a step-by-step practical procedure from the analysis of the current status of the spare parts inventory system to advanced service level analysis by virtue of simulation-optimization technique for a real-world case study associated with a seaport. The remarkable variety and immense diversity on one hand, and extreme complexities not only in consumption patterns but in the supply of spare parts in an international port with technically advance port operator machinery, on the other hand, have convinced the managers to deal with this issue in a structural framework. The huge available data require cleaning and classification to properly process them and derive reorder point (ROP) estimation, reorder quantity (ROQ) estimation, and associated service level analysis. Finally, from 247000 items used in 9 years long, 1416 inventory items are elected as a result of ABC analysis integrating with the Analytic Hierarchy Process (AHP), which led to the main items that need to be kept under strict inventory control. The ROPs and the pertinent quantities are simulated by Arena software for all the main items, each of which took approximately 30 minutes run-time on a personal computer to determine near-optimal estimations.

## I. Introduction

An international operating port with no intermission, high imposed cost due to stops causing by lack of equipment and improper inventory system to maintain the essential functionality of machines, challenging inventory monitoring owing to an immense diversity of items and the varied significance level of items increased the awareness of managers about the service level analysis of the inventory items of Shahid Rajaee Container Terminal, Iran. This scientific and practical approach offers a step by step procedure to be taken by similar medium- and large-sized industries as a guideline to successfully establish an effective inventory system as following three stages:

*1) Stage one:*
  *a) Identifying the inventory items (variety, and consumption rate)*
  *b) ABC analysis (combining with the AHP method)*
  *c) Scoring inventory items in terms of importance and viability to apply advance inventory control methods*
  *d) Selecting the main items according to their score*

*2) Stage two*
  *a) Identifying the demand distribution function of the main items*
  *b) Identifying the distribution function of their lead times*
  *c) Identifying the inventory control costs (including holding, order and shortage cost) for the main items*

*3) Stage three*
  *a) Identifying the expenses for procurement and supply of the main items*
  *b) Simulation of the inventory control system and its analysis*

*c) Determining the optimal inventory policy for the main items (ROP and its quantity)*

*d) Service level analysis*

Therefore, this original research, in one point of view, walks the enthusiastic readers through each step and resolves technical halts s/he might encounter in inventory monitoring. Alternatively, the novel approach comprehends several contributions that distinguish this study from its counterparts, (a) the way it deals with extraordinarily huge data set requires to be cleaned and processed with regard to the importance of each item, (b) embedded simulation-optimization technique and employed Rolette Wheel concept to derive the demand quantity for the main inventory items, (c) not only the real data set but also the technical expertise verification of the final simulation model and catered analysis are valuable assets to this academic and practical research.

In the following, first, we provide relevant and state-of-the-art literature thus far. In section III, the problem statement is described. Solution methodology encompassing the mentioned three steps is in section IV. In the next section, service level analysis is individually furnished as one of the advanced technical analysis tools. Finally, to conclude the study, managerial insights with a concentration on inventory management is suggested.

## II. Literature

Inventory items (for instance: raw materials, semi-finished products, finished products, and spare parts) are among the assets that require acute control and efficient management. There are compelling grounds to store inventories, (a) to prevent interruptions in production operations, (b) to prevent bottlenecks in operation, (c) to timely produce and deliver, accompanying with quality customer services, (d) to operate in the same level, working-station-wise, and (e) to protect against shortage and price fluctuation in raw materials. Therefore, many companies are ought to store inventory to swiftly deal with these prevailing conditions and maintain not only an acceptable service level but also minimal inventory cost. A study of well-established companies' balance sheets shows that about 15% to 35% of companies' capital comprehends inventories. It also demonstrates that approximately 17% to 24% of the monetary value of inventories is spent on holding costs, furthermore, the inventory-to-sales ratio is about 12% to 20% and the inventory-to-asset ratio is approximately 16% to 30% [1].

Since the paramount importance of inventory control is clearly articulated, the concerned case study of this research is not an exception bearing in mind the 24700 items, likewise. One definition of service level that could be modified for our research is obtained from [2], decently defined as "the probability to meet demand while an order is placed". The main focus of the research, however, to examine the effect of safety levels on $\delta$-service level which refers to "the probability to meet a demand with a substitute item while an order is placed". One prior study that stands out among the scholarly works, in the sense of similarities, is conducted by [3] on the subject of fashion supply chain inventory. The optimal inventory model is investigated while the service level and lead time are controllable. The authors proposed two inventory models responding to buyer's and vendor's inventory behavior. In our approach, however, the optimal inventory model is gained by simulation-based optimization modeling considering that the lead time is dictated by the vendors historical data, and a range of service level studied to cater DM with options associated with incurred cost on the entire system of inventory. Although the definition of service level seems wide-range, Candas and Kutanoglu [4] employed customer-based service levels for their inventory model integrated with a location-allocation problem. Transchel and Hansen [5] proposed an inventory model considering service level constraints and uncertainty in lead time while the cornerstone is to take into account the perishability of products tackled by simulation-based optimization method. As elucidated, the perishability of products refers to either having finite shelf life or physical quality decay over time.

The simulation approach is an extensively employed method used to model the complex inventory systems and diverse inventory policies. In one study that combines simulation analysis with big data in retail environment proposed by [6]. The provided case study observed an automated refreshment system in Japan with 3000 items that continually updated by real POS data. The simulation-based algorithm led to a lower inventory level and maintained the service level at the same time. Reference [7] also proposed a simulation-based optimization method called "Sample Average Approximation" in which the service-level is treated as a constraint. Kim and Jeong [8] combined a time series analysis – called ARIMA in order to predict the demand in accordance with the EOQ model – and a simulation model to measure the effectiveness of the model. Along with the previous paper, [9] employed a simulation-based model to determine the safety stock level using Arena software. In their study, they selected only three products to investigate the relationship between the safety stock level and service level. One novel research paper using VenSim software [10] proposed a system dynamic simulation approach to improve inventory performance in the consumer products trade business unit. According to the obtained results, proposed modifications resulted in forty-three thousand dollars saving of operational cost per year. To conclude this paragraph on the simulation technique to deal with the inventory problems, please refer to these recent papers as well, [11]–[13].

As clarified earlier, one segment of this research and the scientific trend is to use varied ABC methods to classify inventory items. Eraslan and Tansel IC [14] proposed an improved decision support system (IDSS) for inventory classification utilizing the ABC method. In the developed software, the proper ABC classification model is selected among five approaches, for instance, the Analytical Hierarchy Process (AHP). One of the close research papers to ours is offered by [15] in which the authors hired an AHP model to classify the spare parts inventory management for the aviation industry. The case study is an aircraft maintenance site in Indonesia intended to reduce unnecessary site stop times. The authors claim that after the validation phase, the model proved to be accurate and rapid in response. In addition to the case studies observed in the literature, one is offered by [16]. In this study with real-world data, an inventory model is provided for the chassis parts at the US container ports. The mentioned study provides a mathematical model that could be considered as a decision support system for the planning of neutral chassis.

### III. PROBLEM STATEMENT

Inventory management plays a critical role in any business, whether manufacturing, or services, and with proper control, one significant step is taken to balance the flow of operations. Inventory control is a common delicate matter that all kinds of companies need to cope with. Moreover, how to handle the mentioned matter is the difference between successful companies and unfortunate ones. In developing countries, the capital stored in the form of inventories is usually higher in comparison with developed countries [17]. Therefore, an accurate estimation of ROP and the quantity of order prevent accumulating and freezing an enormous amount of money in form of inventory items, or system breakdown in terms of disruption in manufacturing or for services rendered, otherwise.

Here, inventory cost entails three well-known parts namely, holding cost, order cost, and shortage cost which is considered in the model. In the following, Fig. 1, a schematic design of the ROP model is demonstrated.

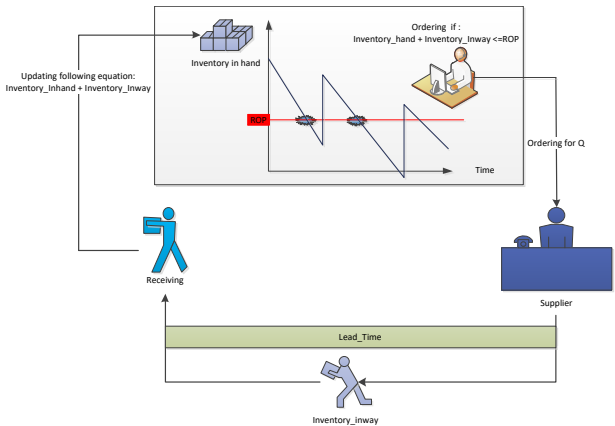

Fig. 1. Conceptual model

### IV. METHODOLOGY

In general, this research has the following features (a) it is categorized into a practical study with its technical contribution; (b) in terms of data collection method, it is descriptive; (c) in terms of analysis point of view, it is a real-world case study; (d) in terms of solution approach, it is quantitative-analytical (Operations Research) which has a foundation on mathematical modeling.

#### A. Step 1: ABC

By properly controlling inventory and planning, you can take steps to balance the flow of operations and reduce the overall production cost. In inventory, a small number of items usually account for the bulk of the inventory's value, and vice versa, a large number of items have a small monetary value. Therefore, having a single inventory control method does not seem reasonably sound for all of these items. In this successful technical project, the ABC analysis is used to identify items whose inventory control is more effectively beneficial in the long run. In addition to basic criteria such as consumption volume and purchase price, other tangible factors influencing the classification of items are also considered. Pareto's analysis method, which is well celebrated and recognized as ABC classification, is that items are divided into three groups (or more) A, B, and C, as follows: group A: includes goods with a high monetary value (or consumption), which is a small percentage. Group B: includes goods with an average monetary value, which also includes an average percentage. Finally, group C: contains goods with low monetary value, but a high percentage. The result of such an analysis shows that we must strictly control class A, control class B less, and control class C less than the first two classes, Fig. 2.

The purpose of this method is to employ the coefficient of importance or necessity of items, which is obtained by questioner and interview with experts and managers of the company which led us to determine the true value of items not only regarding their monetary values, merely. Finally, the relative value of all items is prepared and based on that, a prioritization is determined for them. The steps for inventory items classification into high-, medium-, and low-value are as follows:

*a) Make a list of the total annual value of all items stored in the inventory and sort them by the highest total value.*

*b) Assign a degree of value (importance or necessity) for each item (the largest value will be 1)*

*c) Prepare a table of total values*

*d) Calculate the ratio of the total annual value of each item in inventory*

In Fig. 2, the horizontal axis shows the number of inventory items and the vertical axis shows the percentage of the class's value.

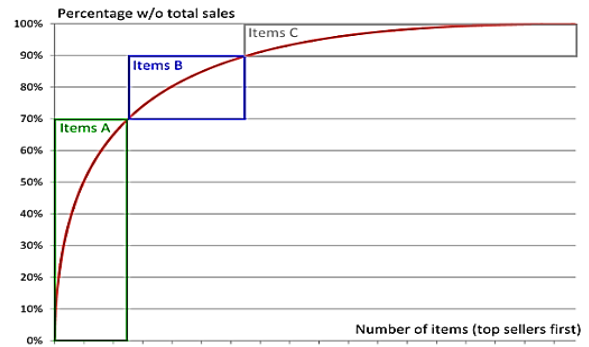

Fig. 2. ABC classification

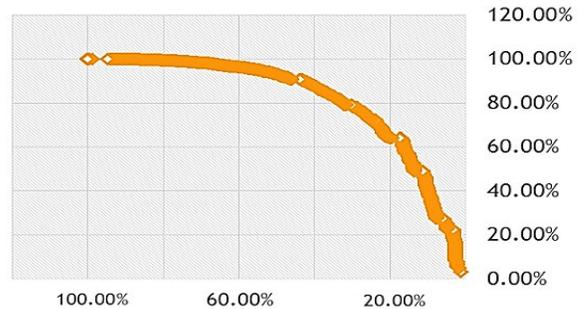

Fig. 3. Inventory value to the inventory level

By considering several ABC analyses (including Fig. 3 and Fig. 4) and according to the Pareto chart, the consumption value to inventory level (Fig. 4) was selected as the best method of Pareto analysis of inventory items. On the other hand, preliminary studies have shown that many of the items in the warehouses are stagnant or the available data are not accurate enough.

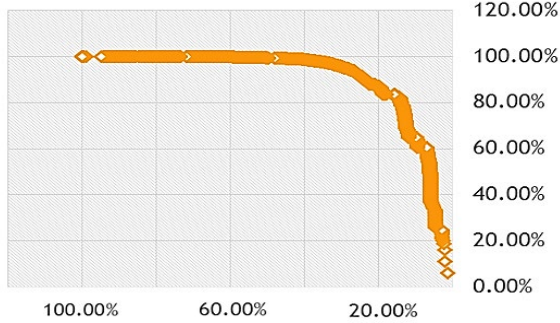

Fig. 4. Consumption value to the inventory level

### B. Forecasting

The most common possible models, which are commonly used from two different approaches to predicting inventory systems, are Poisson and Exponential distribution functions. In the Poisson probability mass function, λ represents the average number of Kanban orders in the time interval [0, t].

$$P[N(t) = n] = \frac{e^{-\lambda t}(\lambda t)^n}{n!}, \quad t \geq 0, \quad n = 1, 2, \ldots \quad (1)$$

When events occur over time randomly and based on a Poisson process with a λ rate, then the time required for an event to occur will have a 1/λ parameter of the exponential Probability Density Function (PDF).

$$f(x) = \frac{1}{\lambda} e^{\frac{-x}{\lambda}}, \quad x > 0, \quad \lambda > 0 \quad (2)$$

By identifying the unknown parameters introduced in the above equations, we can use the probability density functions and various software, such as Arena, Minitab in order to predict the demand and waiting time for inventory items in the future periods.

### C. Step 2: AHP

One of the most effective decision-making techniques is the AHP method proposed by Thomas L. Saati in the 1970s. This technique is based on pairwise comparisons and allows managers to study different scenarios. The AHP has been received warm welcomes by various managers and users due to its simplicity yet comprehensive nature. This technique makes it possible to formulate the problem hierarchically, and besides, it is possible to consider different quantitative and qualitative criteria in the problem. This process involves a variety of options in decision making and allows for the analysis of sensitivity for criteria and sub-criteria. The method is also based on pairwise comparisons that facilitate comparison and calculation. Also, it can show the degree of compatibility of the decision, which is one of the excellent advantages of this technique in multi-criteria decision making. Saati has stated the following four principles as the principles of the hierarchical analysis process and has based all calculation rules on these principles. These principles include: (a) Reciprocal Condition, (b) Homogeneity, (c) Dependency, (4) Expectations.

*1) Weights calculation*

The AHP analyzes complex problems, transforms them into simple forms, and solves them. The following steps must be taken to resolve an issue or decision [18]:

*a) Model the problem in a hierarchical format encompassing the objective, alternatives, and criteria for evaluating the alternatives.*

*b) Set priorities among the elements of the hierarchy by making a judgment in a row based on pair-wise comparisons of the factors.*

*c) Combine the series of judgment to conclude a set of priorities for the hierarchy.*

*d) Control the consistency of the judgment according to the four mentioned principles.*

*e) Eventually, reach a final decision.*

The calculation of weight in the hierarchical analysis process is discussed in two separate sections: (a) relative weight, (b) absolute weight. The relative weight is obtained from the pairwise comparison matrix, while the absolute weight is the final rank of each option, which is calculated from the combination of relative weights.

*a) Relative weight*

In the AHP, the elements are first compared in pairs and the matrix of pairwise comparisons is formed. Then, using this matrix, the relative weight of the elements is calculated. In general, a pair-wise comparison matrix, $A$, is shown as follows, in which the $a_{ij}$ is the preference of the element $i$ over the element $j$, therefore, the weight of the element, $w_i$, is obtained according to the value of $a_{ij}$.

$$A = \begin{bmatrix} a_{11} & a_{12} & \ldots & a_{1n} \\ a_{21} & a_{22} & \ldots & a_{2n} \\ \vdots & \vdots & \ddots & \vdots \\ a_{n1} & a_{n2} & \ldots & a_{nn} \end{bmatrix} \quad (3)$$

Any pairwise matrix, $A$, may be consistent or inconsistent. When this matrix is consistent, the calculation of the weights ($w_i$) is simple and is obtained by normalizing the elements of each column. However, in a case where the matrix is inconsistent, it is not easy to calculate the weights and there are four main methods for obtaining it, which are: (a) least-squares method, (b) logarithmic least squares method, (c) eigenvector method, and (d) approximation methods. It should be noted that since the first three methods have heavy computations, some approximate methods have been proposed that are less accurate and have less and simpler calculations. These methods are mainly approximations of the eigenvector method that correct the calculations with different accuracy.

*b) Absolute weight*

The final weight of each option is obtained from the sum of the multiplication of the value of the criteria and the weight of the option.

## 2) Consistency calculation

As mentioned earlier, a matrix may or may not be consistent. In a consistent matrix, the calculation of weight is simple and is achieved by normalizing a single column, while several methods have been mentioned for calculating weights in the inconsistent matrix. Calculating the level of inconsistency is very important because it shows how much confidence can be given to priorities from comparisons. In general, it can be said that the acceptable level of the inconsistency of a matrix depends on the decision-maker, but Saati presented the number 0.1 as an acceptable limit and believed that if the level of inconsistency is more than 0.1, it is better to reconsider the judgment.

### a) Consistent matrix

If there are $n$ criteria namely $c_1, c_2, ..., c_n$, and the pair-wise comparison matrix is as follows:

$$A = [a_{ij}]; \ i, j = 1, 2, ..., n \quad (4)$$

Where $a_{ij}$ shows the preference of the element $c_i$ over $c_j$ if the following holds:

$$a_{ik} \times a_{kj} = a_{ij}; \ i, j, k = 1, 2, ..., n \quad (5)$$

Then, $A$ is a consistent matrix.

### b) Inconsistent matrix

In this section, we want to know if the matrix of the pair-wise comparison is inconsistent, what is the degree of inconsistency of the matrix and how do we measure it? Before stating the criteria for measuring matrix inconsistency, the following are three important theorems about any matrix of pair-wise comparison:

Theorem 1: if $\lambda_1, \lambda_2, ..., \lambda_n$ are the eigenvalues of matrix $A$, the sum of all eigenvalues equals $n$.

$$\sum_{i=1}^{n} \lambda_i = n \quad (6)$$

Theorem 2: The largest eigenvalue ($\lambda_{max}$) is always greater than or equal to $n$ (in which case some $\lambda$ will be negative):

$$\lambda_{max} \geq n \quad (7)$$

Theorem 3: If the matrix elements deviate a small distance from the consistency state, the eigenvalues will also deviate a small distance from their consistency state.

$$A \cdot w = \lambda \cdot w \quad (8)$$

Where $\lambda$ and $w$ are the eigenvalues and the eigenvector of matrix $A$, respectively. When matrix $A$ is consistent, a special value is equal to $n$ (the largest eigenvalue) and the rest is zero. Therefore, in this case, the following could be stated:

$$A \cdot w = n \cdot w \quad (9)$$

If the matrix of A is inconsistent, according to Therem 3, $\lambda_{max}$ moves slightly away from $n$, so we have:

$$A \cdot w = \lambda_{max} \cdot w \quad (10)$$

The reason to use $\lambda_{max}$ according to Theorem 3 is that it will be the shortest distance from $n$. Since $\lambda_{max}$ is always greater than or equal to $n$, and if the matrix deviates slightly from the consistency state, $\lambda_{max}$ will deviate slightly from $n$, so the difference between $\lambda_{max}$ and $n$ ($n - \lambda_{max}$) depends on the value of $n$, and to eliminate this dependency, an index can be calculated as the following. The index is the definition of the Inconsistency Index (II):

$$II = \frac{\lambda_{max} - n}{n - 1} \quad (11)$$

The values of the inconsistency index (I.I.) are calculated for matrices whose numbers are completely random, and it is called the Random Matrix Inconsistency Index (R.M.I.I.), the values for the following $n$ matrices are as follows:

TABLE I. RANDOM MATRIX INCONSISTENCY INDEX (R.M.I.I.)

| n | 1 | 2 | 3 | 4 | 5 | 6 | 7 | 8 | 9 |
|---|---|---|---|---|---|---|---|---|---|
| R.M.I.I. | 0 | 0 | 0.58 | 0.9 | 1.12 | 1.24 | 1.32 | 1.41 | 1.45 |

For each matrix, the result of dividing the inconsistency index (I.I.) by the random matrix inconsistency index (R.M.I.I.) is a proper criterion for making a judgment about inconsistency, which is called the inconsistency rate (I.R.). If this number is less than or equal to 0.1, the system's inconsistency is acceptable, otherwise, the process must be reconsidered. Finally, according to the mentioned procedure, the AHP method is performed with the assistance of Expert Choice software. In the following, the final five criteria – among twenty-seven well-known ones in literature studies – to classify inventory items were selected by experts' opinion:

(a) Critical degree: Critical goods or materials are those items that are considered critical by the company for reasons related to their consumption or purchase process, and its deficiency can be detrimental.

(b) Item Consumption: There are different parts of the site where the items are consumed.

(c) Lead time: The time between ordering and receiving the goods in stock.

(d) Availability: Possibility to provide, purchase and transfer items quickly to the warehouse due to the abundance of items or the availability of supply contracts.

(e) Inventory turnover: In accounting, it is a ratio, which indicates that the company's inventory of items and materials, in a certain period of time (such as in one fiscal year), has been consumed or sold several times and replaced.

Using Expert Choice software, we obtained the weight of each criterion. The weight values for the criteria are in the following table:

TABLE II. FINAL CRITERIA WEIGHTS

| Criteria | (a) | (b) | (c) | (d) | (e) |
|---|---|---|---|---|---|
| Weight | 0.52 | 0.15 | 0.14 | 0.12 | 0.7 |

### D. Calculate the qualitative and multivariate values of items using the AHP concept

To classify items using the AHP method, we first normalize the values of the items in each criterion. The normalized value of each item for each criterion was obtained based on the common procedure in the AHP method linearly and by dividing each of the values by the sum of the corresponding standard columns. The method of calculating the relative rank of each item can be expressed as follows:

$$R_i = \sum_{j=1}^{5} W_j \cdot V_{ij} \qquad (12)$$

Where:

$W_j$ is the relative weight of criterion $j$, and

$V_{ij}$ is the normalized value for item $i$ regarding criterion $j$.

After calculating the qualitative value for all main items, the consumption value was obtained by multiplying the consumption amount by the price. Due to the fact that the qualitative values obtained for the items were normalized and to combine quantitative and qualitative values, these quantitative values of consumption were also normalized by adding any value to the total quantitative value of the goods. In order to combine quantitative and qualitative values for each item and to calculate the combined value, it was necessary to consider a weight for each of these two types of values. Due to the involvement of several criteria in calculating the qualitative value, an attempt was made to consider a higher weight for the qualitative value against the quantitative value of the sole quantitative criterion. After performing different scenarios and finally according to the quality of the Pareto diagram obtained for each scenario, the ratio of 6 to 1 was selected for qualitative value compared to quantitative value. Therefore, how to calculate the combined value of each product can be expressed as follows:

$$G_i = \frac{6}{7} R_i + \frac{1}{7} K_i \qquad (13)$$

Where:

$R_i$ is the qualitative value of several criteria obtained for item $i$

$K_i$ is the quantitative value of the amount of consumption obtained for item $i$

$G_i$ is the combined value obtained for item $i$

After calculating the combined value for all items, the initial classification of goods was done by the ABC classification method. In this classification, the Pareto diagram of items to determine items in classes A, B, and C were obtained by comparing the value of inventory with the number of items in the inventory, which is as follows, Fig. 5. The vertical axis is the cumulative percentage of integrated values and the horizontal axis is the cumulative percentage of inventory. Finally, 1416 items were selected for simulation analysis with Arena software.

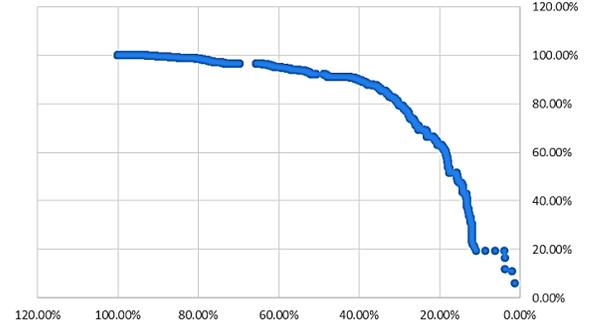

Fig. 5. Integrated ABC classification with AHP

### E. Step 3: Simulation

EXTRAPOLATION

The first step to derive the required information for initializing simulation is to conduct data gathering. The most important source of information and analytical basis for identifying the distribution function of demand and lead time is the historical information that can be extracted from the existing inventory system and documents or through the opinion of the company's experts. In addition to historical information, the company's development plans and estimation of future trends can also be considered.

The first analysis that needs to be done to model and simulate 1416 main items of the inventory is to identify their distribution function (fit the probability density function). For this purpose, according to the list of probability density functions available in Arena software, the fit algorithm was implemented in the MATLAB software environment. The reason for using MATLAB software is the higher ability and accuracy of this software in fitting probability density function. After identifying the most appropriate function for each item, the type of function and its parameters are entered into the simulation model in Arena as required inputs. In this project, the Bayesian Information Criterion (BIC) is used to select the best distribution for data of demand. The index is calculated as follows:

$$BIC = k \times \ln(n) - 2 \times \ln(L) \qquad (14)$$

In which, $k$ is the number of parameters to be estimated, $n$ is the number of observations, and $L$ is the maximum value of the correction function for the model. Among the candidate distribution functions, the one with the lowest *BIC* level is the best fit. The procedure used to fit the distributions in this project is in the form of a flowchart shown in Fig. 6.

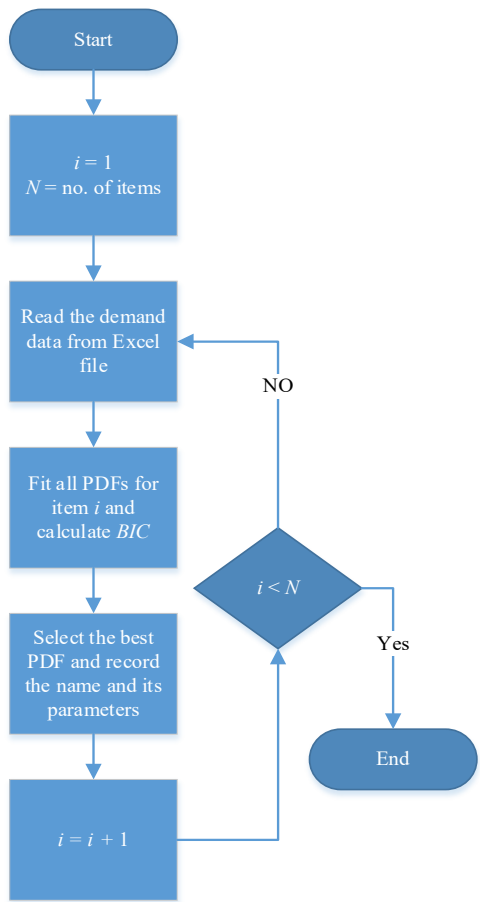

Fig. 6. The procedure used to fit the distribution function

IMPLEMENTATION

In order to establish a simulation model, the following steps must be taken:

1) Determining the boundary of the studied system
2) Making a model in the software (here Arena 14.0)
3) Determining the nature of the model in terms of a finite or a non-finite system (here the inventory control system is classified as a non-finite system [19]).
4) Confirmation and validation of the model implementation in accordance with the performance of the real-world model
5) Discussion and reports

The inventory control model in this software consists of three sub-sections. Moreover, the demand period is annual.

The first part deals with demand, in which the following assumptions are considered:

1) Demand enters the system at the end of each year.
2) The amount of demand per month is proportional to the annual demand for inventory items in the last 9 years.
3) With each entry of demand into the model, the amount of inventory accumulated with demand is examined and questioned, while there are two answers:

   a) The demand can be met by the available inventory (in-hand inventory).
   b) The amount of inventory is not sufficient to fully satisfy the demand, in other words, part or all of the demand cannot be satisfied; in this case, the amount of unsatisfied demand is considered as a shortage.

The second part is related to inventory supply system and inventory monitoring, here the ROP-ROQ method is used. The following assumptions are also used in this section.

1) The inventory level is monitored online and continuously so that after each order arrival, the total inventory level (in-hand inventory) and the coming items (on-way inventory) is less or equal to the ordering point (ROP), a new order is placed.
2) With the occurrence of the order, the system status changes to be ready to monitor the inventory level.
3) When the time for receiving the orders arrives, they are immediately added to the inventory of the warehouse and change the status to the ready-to-use state; that is, the amount of inventory on the way is added to the inventory in hand. And for each order, the corresponding order cost is considered.

The third part; this part has two sub-sections:

1) Recording holding cost:
The holding cost period is considered to be annual, for which purpose a separate part of the model is responsible for updating the holding costs. At the end of each year, the amount of remaining inventory in the warehouse is multiplied by the holding costs unit and recorded.

2) Record all costs, including holding, ordering, and shortage costs:
In order to collect all the costs imposed on the system, there is a section in Arena software that allows the user to collect the costs at the end of the simulation period (end of simulation clock). This section is called the Statistic data module:

GENERAL SIMULATION PROCESS (ARENA) - OPTIMIZATION (MATLAB)

After constructing, reviewing, and approving the inventory system simulation model, in a general process, a simulation-optimization method of this system was performed. This general process, which was simulated in Arena software and optimized in MATLAB software, accompanied by the VBA interface, and illustrated as follows.

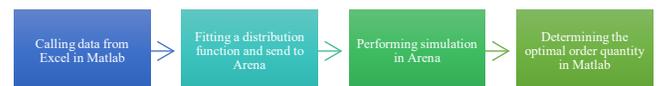

Fig. 7. Simulation-optimization procedure

*Roulette Wheel*

To simulate the inventory model, it is necessary to predict the amount of consumption in different periods, as well as extracting the history of consuming items without considering overhaul from the database and get the correct pattern of demand behavior. The prediction model is as follows (Fig. 8):

Using annual consumption data of spare parts, we form a distribution function. The distribution function indicates the chance of occurrence of each of the consumption ranges. For example, in Fig. 8, section A, the probability that the amount of spare parts used in the first interval is more likely than the other intervals (the probability of occurrence is 4/11, while for the other intervals it is 1/11, 2/11, 3/11, and 1/11 respectively). Now, using these possibilities, we use the Roulette Wheel technique. The Roulette Wheel is such that the more probable the range, the larger the area of the wheel. As shown in Fig. 8, part B, the green part indicates the highest chance of occurrence. Now, if we rotate the Roulette Wheel, the indicator will be selected in each interval that stands, and the amount of the annual consumption prediction in that interval will be determined.

Because the simulation accuracy is higher per month, the annual consumption should be converted to monthly. This is done using a uniform distribution so that each month of the year has a chance to be equal in consumption. Also, the amount of consumption in months should be such that it is equal to the total annual forecast. As shown in Fig. 8, sections C is first randomly selected for the first, sixth, and tenth months, and then the corresponding consumption values (1, 2, 1, respectively) are assigned to the accident.

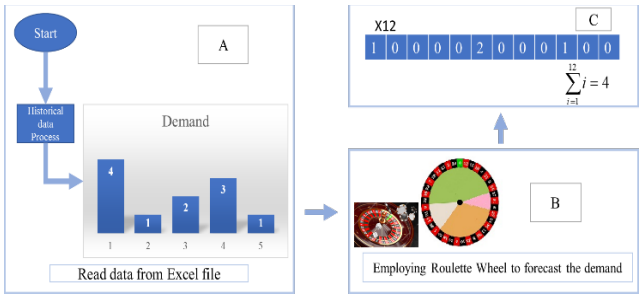

Fig. 8. Rolette wheel

## V. SERVICE LEVEL ANALYSIS

It should be noted that the service level means the possible percentage of meeting the demand for items, and it is a function of ROP value. On the other hand, the specified ROP value determines the amount of capital expenditure according to the holing cost of items. By service level analysis, we provide the fluctuation in total inventory cost. Fundamentally, by changes in service level and inventory level, an analysis of cost is provided.

<u>Calculation logic</u>: The level of service measures the probability of a shortage. Therefore, the order point must be defined in such a way that it can respond to the demand after the order is issued until the items are received, that is, lead time. Therefore, the distribution function of the occurrence probability of demand at the waiting time must be recognized to obtain the probability of shortage according to the re-order point. After obtaining the demand distribution at the waiting time, according to the desired service level, a re-order point is computed. In the following, the computation steps for the re-order point is explained:

*1) Recognize the demand distribution based on historical data.*

*2) Demand distribution is obtained for the lead time. Mean and variance are obtained for the lead time according to the following equations (in the following equation, instead of a, we plug in the lead time):*

$$E(aX) = a(E(x));\ Var(ax) = a^2 Var(x) \qquad (15)$$

*3) According to the obtained mean and variance, the distribution parameters are corrected and the demand distribution is obtained for the lead time.*

*4) Depending on the service level of α, the re-order point is obtained through the following relationship:*

$$ROP = F^{-1}(\alpha) \qquad (16)$$

Based on the above procedure and in order to draw the relationship between service level and inventory costs, the total inventory cost for the following service levels 50%, 70%, 80%, 90%, 95%, 98%, 99% and 99.99% calculated and the obtained results are shown in Fig. 9. As can be observed in Fig. 9, the higher the service level, the higher the overall inventory cost will be. In other words, the additional investment on the inventory system while changing the service level from 0.8 to 0.9 is much higher than adapting the service level from 0.6 to 0.7.

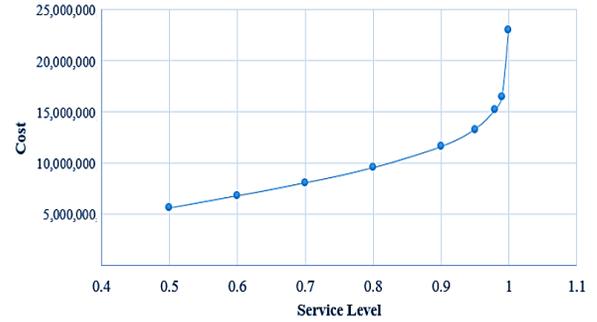

Fig. 9. Service level – cost analysis

## VI. CONCLUSION

In this paper, we aimed at proposing a technical and novel approach to properly monitor the inventory system for the main items. Firstly, an ABC classification is offered, then, with the assistant of the AHP methodology, an integrated approach of qualitative and quantitative methods led us to an applicable ABC classification. While the classification is made, the class A consisting of 1416 items were identified as the main items of the inventory worthy of strict control. Finally, a simulation optimization method determines the best reorder point and the quantity of each order in the recognized inventory policy reorder-point, reorder-quantity model. The following suggestions are provided which is founded on practical experience and close observation of the system:

- Implementing new parts supply methods, such as vendor managed inventory (VMI) to reduce or eliminate the inventory of low-consumption and stagnant items.
- Diagnosis of the processes of supplying low-consumption and stagnant parts.
- Considering an effective integration of maintenance and repair processes pertinent to the inventory items.

- It is suggested that appropriate key performance indicators (KPIs) need to be calculated from the perspective of efficiency and optimization of operating costs, such as inventory cost in detail, inventory turnover.
- In the case of non-stagnant items in class B of Pareto classification, mixed inventory policies are proposed such as periodic inspections.

To address future pathways for this technical paper, we would like to suggest two approaches. Firstly, fuzzy techniques could be a perfect complement to the ABC method since it allows considering quantitative and qualitative criteria simultaneously. The fundamentals could be actively sought in [20], [21]. Secondly, heuristics and meta-heuristics algorithms could be an alternative to combine ABC classification with, for instance, consideration of hybrid ABC-GA approach for deteriorating items in inventory by [22], and stochastic optimization by [23], [24].

ACKNOWLEDGMENT

We would like to cherish the moment and sincerely thank the managers of Shahid Rajaee Container Terminal in Iran who were ardently support this non-profit academic project by sharing insensitive data with us. This paper is the fruit of the final academic project involving collaboration with industry at the University of Shahid Beheshti in Tehran, Iran.